\documentclass[preprints,article,accept,moreauthors]{Definitions/mdpi} 

\firstpage{1} 
\pubvolume{1}
\issuenum{1}
\articlenumber{0}
\pubyear{2022}
\copyrightyear{2022}

\datereceived{} 

\dateaccepted{} 
\datepublished{} 

\hreflink{https://doi.org/}

\usepackage{epsfig}

\usepackage{epstopdf}
\usepackage{bm}
\usepackage{color} 

\Title{Complex phase-fluctuation effects correlated with granularity in superconducting NbN nanofilms}

\TitleCitation{Title}

\Author{Meenakshi Sharma $^{1,\ddagger}$\orcidA{}, Manju Singh $^{2}$, Rajib K. Rakshit $^{2}$, Surinder P. Singh $^{3}$, Matteo Fretto $^{4}$, Natascia De Leo $^{4}$, Andrea Perali $^{5}$\orcidE{}, Nicola Pinto $^{1,4\ddagger}$\orcidF{}}

\AuthorNames{Firstname Lastname, Firstname Lastname and Firstname Lastname}

\AuthorCitation{Lastname, F.; Lastname, F.; Lastname, F.}

\address{%
$^{1}$ \quad School of Science and Technology, University of Camerino, Camerino, Italy\\
$^{2}$ \quad CNR-Institute for the study of Nanostructured Materials, Bologna, Italy \\
$^{3}$ \quad National Physical Laboratory, CSIR, New Delhi, India\\
$^{4}$ \quad Advanced Materials Metrology and Life Science Division, INRiM, Torino, Italy\\
$^{5}$ \quad School of Pharmacy, Physics Unit, University of Camerino, Camerino, Italy}
\corres{Correspondence: nicola.pinto@unicam.it}

\secondnote{These authors contributed equally to this work.}

\abstract{Superconducting nanofilms are tunable systems that can host a 3D-2D dimensional crossover, leading to the Berezinskii-Kosterlitz-Thouless (BKT) superconducting transition approaching the 2D regime. Reducing further the dimensionality, from 2D to quasi-1D, superconducting nanostructures with disorder can generate quantum and thermal phase slips (PS) of the order parameter. Both BKT and PS are complex phase fluctuation phenomena of difficult experimental detection. Here, we have characterized superconducting NbN nanofilms thinner than 15 nm, on different substrates, by temperature dependent resistivity and current-voltage (I-V) characteristics. Our  measurements have evidenced clear features related to the emergence of BKT transition and PS events. The contemporary observation in the same system of BKT transition and PS events and their tunable evolution in temperature and thickness, has been explained as due to the nano-conducting paths forming in a granular NbN system. In one of the investigated samples we have been able to trace and characterize the continuous evolution in temperature from quantum to thermal PS. Our analysis has established that the detected complex phase phenomena are strongly related to the interplay between the typical size of the nano-conductive paths and the superconducting coherence length.}

\keyword{NbN; ultrathin films; BKT transition; phase slips; granular superconductivity} 

\begin{document}

\section{Introduction}
Effects related to thermal and quantum fluctuations in low-dimensional superconductors, as phase slips  \cite{Bezryadin2000,Bezryadin2008,Zhao2016,Lehtinen2012,Baumans2016}, quantum criticality \cite{Kim2018}, superconductor-insulator transition \cite{carbillet2016confinement}, quantum phase transitions \cite{Mason1999,Breznay2017} have been studying for several decades. These quantum and many-body effects are controlled by several film properties as spatial dimensions, electronic disorder and structural inhomogeneities \cite{gajar2019substrate,bell2007one}.\\ 
In recent years, the scientific interest has been focused to quasi 2D and 1D systems, where the presence of resistive states close to the superconducting transition temperature $T_c$ have been found to produce detectable effects in the transport properties \cite{bell2007one,gajar2019substrate}.\\ 
In such systems, the emerging resistive states are a fundamental phenomenon, which involves understanding of advanced concepts as topological excitations, phase disorder, and interplay between different length scales of the superconducting state. These have important applications in the field of quantum technologies, ultrasonic detectors of radiation, single photon detector and nanocalorimeters \cite{bell2007one}.\\
Efforts have been carrying out to study low-dimensional systems theoretically while, from the experimental side, attention has been focusing on thin films in which  a crossover from 3D to 2D Berezinskii-Kosterlitz-Thouless (BKT) transition \cite{Bartolf2010,Sidorova2020,Koushik2013,Venditti2019} occurs, lowering the film thickness to few nanometers as, for instance, in NbN \cite{Koushik2013, Berezinskii1971,Berezinskii1972,Kosterlitz1973,Kosterlitz1974}. \\
Superconducting to normal state transition in a 2D-XY model was introduced so far to explain the formation of thermally excited vortex-anti-vortex pairs (VAP) in ultrathin films, even in absence of an applied magnetic field \cite{Bartolf2010}, resulting into the BKT topological phase transition \cite{bartolf2010current,yong2013robustness}. The nature of a BKT transition is completely different from the standard second order phase transition given by the Landau paradigm. 
It is driven by the binding of topological excitations without any symmetry breaking associated with the onset of the order parameter \cite{giachetti2021berezinskii}.\\
A BKT-like transition is expected to occur even in dirty superconductors, at a reduced film thickness $d$, under some physical constraints. A required condition is that the Pearl length $\Lambda = 2\lambda^2/d$ ($\lambda$ being the London penetration depth), exceeds the sample size with negligible screening effects due to charged supercurrents \cite{Beasley1979}. 
Furthermore, $\lambda \gg w$ or $d \ll \xi_{GL}$ ($\xi_{GL}$ being the Gintzburg-Landau coherence length) must be fulfilled in order to detect BKT effects, while some experiments have reported BKT transition also outside the theoretically established limits \cite{chu2004phase}.\\
To observe this remarkable phenomenon in real systems, two approaches have been explored. In the first, the BKT theory predicts a universal jump in the film superfluid stifness, $n_s$, at the characteristic temperature $T_{BKT} < T_{MF}$, $T_{MF}$ being the mean-field superconducting transition temperature. This jump is related to the VAP binding through a logarithmic interaction potential between free vortices. Sourced current can break bound pairs, producing free vortices, inducing non linear effects in I-V curves \cite{chu2004phase}. For the second approach, transition is observed in the correlation length, which diverges exponentially at $T_{BKT}$, in contrast to the power-law dependence expected within the Ginzburg-Landau (GL) theory \cite{benfatto2000kosterlitz,mondal2011role}.\\
Reducing further film size, effect of fluctuations can result into the formation of multiple resistive states in I-V curves, at both low and high $T$ range. These intermediate states form due to the change in phase of the order parameter by 2$\pi$ and they will result in a discontinuous voltage jump, forming phase slips. These PS are characteristic of quasi 1D system, which form by a river of fast moving vortices (kinematic vortices) driven by the topological excitations, annihilating in the middle of the sample \cite{paradiso2019phase,Bartolf2010}.
However, several superconductors have been explored experimentally to study such topological effects such as NbSe$_2$ \cite{paradiso2019phase}, NbN \cite{Bartolf2010}, Nb$_2$N \cite{Gajar2019}, Nb \cite{rezaev2020topological} and many more.\\
In particular, NbN is a known and well studied material, belonging to the family of strongly coupled type-II superconductors, potentially interesting for several technological applications, due to its relatively high value of the bulk $T_c$ ($\simeq 16$ K), while the small coherence length, $\xi$ ($\approx 4$ nm) requires fabrication of extremely thin films (few nanometers of thickness) with a fine control of their properties to achieve the 2D superconducting regime \cite{alfonso2010influence}.\\
In this work, a detailed experimental study has been carried out about the electrical properties of superconducting NbN ultra-thin films, aimed to investigate the crossover regime from a quasi 2D BKT phenomenon to a quasi 1D PS mechanism. Outcomes evidenced the presence of large fluctuation effects mostly close to $T_c$, with an exponential decrease at lower temperatures.
In one case, the freezing of thermal fluctuations at the lowest temperatures has given rise to a quantum phase slip (QPS) phenomenon, while in the other case two distinct resistive transitions, below $T_c$, have been detected for one of the thinnest NbN films, suggesting the unexpected coexistence of BKT transition and phase-slip phenomena at the dimensional crossover from 2D to 1D.
Therefore, findings on superconducting to resistive-state transition features in investigated thin films of NbN, have been questioning if their origin is due to thermal fluctuations (e.g., thermal PS), quantum fluctuations (e.g., QPS) \cite{delacour2012quantum} or to a proximity effect mechanism among coupled nano-sized superconducting grains \cite{Gajar2019}.
Our study suggests that the thickness threshold where quantum phase fluctuation effects can start to appear is not yet clearly defined, since BKT transition or QPS phenomena have been detected even in NbN films nominally 10 nm thick. Hence, specific conditions to detect BKT transitions and PS events in quasi-2D systems, especially those being granular in nature, deserve to be further studied both theoretically and experimentally.

\section{Materials and Methods}
\subsection{Deposition}
NbN films of nominal thickness of 5 nm, 10 nm and 15 nm have been deposited on several substrates as MgO, Al$_2$O$_3$ and SiO$_2$ (see Table~\ref{tab1}), by using DC magnetron Sputtering. The optimized deposition rate was $\simeq$ 0.4 nm/s, at a substrate temperature of 600$^\circ C$ and at 200 $W$ of discharge power. The N$_2$/Ar ratio was fixed at 1:7, during the deposition process.\\ 
\subsection{Fabrication of the hall bar geometry}
For the electrical characterization, an 8-contacts Hall bar has been fabricated (see inset of figure~\ref{rho_vs_T}) by using a suitably designed mask (LibreCAD software). 
The Hall bar length has been fixed at 1000 $\mu$m while bar widths of 10 $\mu$m and 50 $\mu$m have been chosen. Hall bar mask fabrication has been carried out by using a direct write laser lithography system, from Heidelberg instruments, While Hall bars have been patterned by optical lithography with a mask aligner from KarlSuss (mod. MJB3). The etching process to define the final NbN Hall bar samples has been carried out using a Reactive Ion Etching from Oxford Instrument (Plasma Pro Cobra 100). This fabrication step has been optimized taking into account the different films thicknesses and different substrates types. The best selectivity for the optical resist used (AZ5214E) was obtained with a mixture of CF4 and Ar with a flux of  90 sccm and 10 sccm respectively at a pressure of 50 mTorr. The ICP power was 500 W and the table RF power was 30 W. All the samples have been cleaned with a light oxygen plasma before the etching to eliminate any organic or lithographic residual. Further, the samples were glued to a copper holder with an high thermal conductivity glue (GE Varnish) in order to have a better temperature control and stability in helium free cryostat. Aluminum wires have been bridge bonded to the pads of the Hall bar and to pads of a small pcb, placed close to the sample (see figure \ref{rho_vs_T}). A scanning electron microscope (SEM) has been used to check the intermediate steps of the fabrication process.\\
\subsection{Electrical characterization} 
Resistivity, $\rho(T)$, and current-voltage (I-V) characteristics have been measured as a function of temperature in a liquid-free He cryostat (Advanced Research System mod. DE210) equipped with two Si diode thermometers (Lakeshore mod. DT-670) one out of two calibrated \cite{Pinto2018,Rezvani2019}. A temperature controller Lakeshore mod. 332, has been used to read the temperature of the uncalibrated thermometer, thermally anchored to the second stage of the cryostat. The film temperature has been measured by using one channel of a double source-meter (Keysight mod. B2912A). The other channel of the instrument, has been earmarked for electrical characterisation of NbN film properties ($\rho$ and I-V) in the 4-contacts geometry, sourcing the current (0.1 $\div$ 100 $\mu$A, typically 1 $\mu$A for $\rho$) and detecting the voltage drop. Either dc or a pulsed mode technique have been used.\\
For I-V characteristics, both DC \cite{daire2005improved} and pulse mode techniques have been adopted. For measurements executed by the pulse mode technique pulse mode technique \cite{inc2004achieving}, a sweep of current pulses of increasing intensities, each of duration 1.1 ms, has been used. For all investigated films, current pulse intensities ranging from tens of $\mu$A to few mA have been used. 
Due to the high thermal inertia of the cryostat, data have been collected without any thermal stabilisation in the whole range of temperatures ($\approx 5\div 300$ K), upon sample cooling. The maximum $T$ change, detected at the lowest $T$ during data acquisition (e.g., $\rho$), has been of the order of 15 mK.
For each data point of the resistivity curve, typically 30 values have been averaged, while a suitably selection of the working parameters of the source-meter has allowed to capture several values during the transition from the superconducting to the normal state. During I-V characteristics, typically 200 points have been collected in few seconds for each curve, while the maximum $T$ variation during each I-V curve acquisition has been $\lesssim$ 50 mK.
\section{Results}
\subsection{Superconducting state properties}
\begin{figure}[H]
\centering
\includegraphics[width= 12 cm]{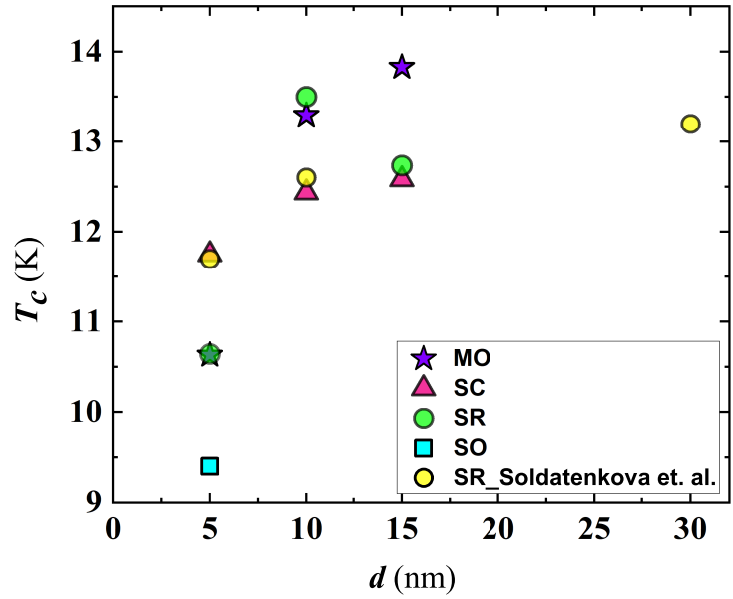}
\caption{Thickness dependence of the superconducting transition temperature of NbN films deposited on MgO (MO), SiO$_2$ (SO) and Al$_2$O$_3$ substrates, both c-cut (SC) and r-cut (SR) types. For comparison, $T_c$ data of films deposited on Al$_2$O$_3$ r-cut substrates by Soldatenkova et {\it al.} \cite{soldatenkova2021normal} have been plotted.}
\label{T_c}
\end{figure}

\begin{figure}[H]
\centering
\includegraphics[width= 12 cm]{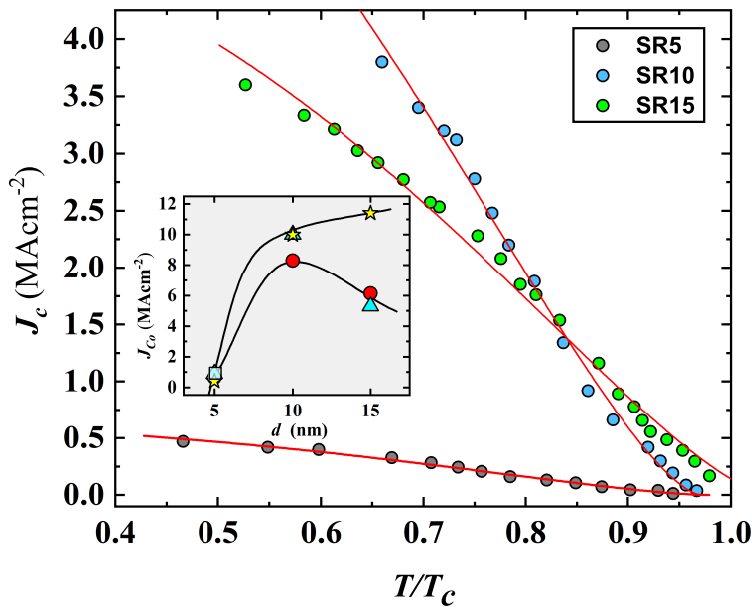}
\caption{Current density as a function of the temperature, normalised to $T_c$, for the whole set of NbN films deposited on the Al$_2$O$_3$ r-cut substrate. Lines are the result of the least squares fitting by the Ginzburg-Landau equation (see ref.~\cite{Pinto2018}). Inset: thickness dependence of the critical current density at zero temperature, $J_{c0}$, for NbN films deposited on Al$_2$O$_3$ r-cut (circles), Al$_2$O$_3$ c-cut (triangles), MgO (stars), SiO$_2$ (square).}
\label{Jc_vs_T}
\end{figure}

NbN film properties have been investigated by resistivity and current-voltage characteristics as a function of the temperature.
\begin{table}[H]
\caption{NbN films properties. Starting from left, columns heading are: film acronym (SC: Al$_2$O$_3$ c-cut; SR: Al$_2$O$_3$ r-cut; MO: MgO; SO: SiO$_2$; the number following the two letters refers to the film thickness in nm units); resistivity value at 15 K; superconducting transition temperature; superconducting transition width; superconducting critical current density at 0 K.}
\label{tab1}
\begin{tabularx}{\textwidth}{CCCCC}
\toprule
\text{Sample}	& \text{$\rho$ ($\Omega$cm)}   & \text{$T_c$ (K)}   & \text{$\Delta T_c$ (K)}   & \text{$J_{c0}$ (MA/cm$^2$)} \\
\midrule
MO5a\textsuperscript{*}	& 8.0$\times10^{-5}$    & 10.072    & 0.08   & 0.40  \\
MO5b\textsuperscript{**}	& 5.8$\times10^{-4}$     & 11.02    & 0.79    & 2.24 \\
MO10	& 1.2$\times10^{-4}$    & 13.29    & 0.27    & 9.98   \\
MO15	& 2.4$\times10^{-4}$    & 13.83   & 0.23    & 11.40   \\
SC5		& 8.0$\times10^{-5}$     & 10.64    & 0.43   & 0.90   \\
SC10	& 2.4$\times10^{-4}$    & 13.50    & 0.40   & 10   \\
SC15	& 1.7$\times10^{-4}$    & 12.73    & 0.24    & 5.29   \\
SR5	\textsuperscript{**}	& 1.1$\times10^{-4}$     & 11.76    & 0.68    & 0.63  \\
SR10	& 1.7$\times10^{-4}$   & 12.43    & 0.30    & 8.3   \\
SR15	& 2.3$\times10^{-4}$    & 12.58    & 0.38    & 6.14   \\
SO5		& 9.3$\times10^{-7}$    & 9.40    & 0.46    & 0.89   \\
\bottomrule
\end{tabularx}
\noindent{\footnotesize{\textsuperscript{*}Both samples MO5a and MO5b belong to the same deposition run but the fabrication process of their  Hall bar  has been carried out by using slightly different parameters.}}\\
\noindent{\footnotesize{\textsuperscript{**}The Hall bar width of this film is 10 $\mu$m.}}
\label{Table_Table 1}
\end{table}
At the superconducting (SC) transition, investigated films show a resistivity jump spanning from $\approx$2 to $\approx$5 orders of magnitude, while $T_c$ depend on the film thickness and substrate type, and its value rapidly decreases at $d\leq 10$ nm (see Table~\ref{tab1}).\\ 
\begin{figure}[H]
\centering
\includegraphics[width= 12 cm]{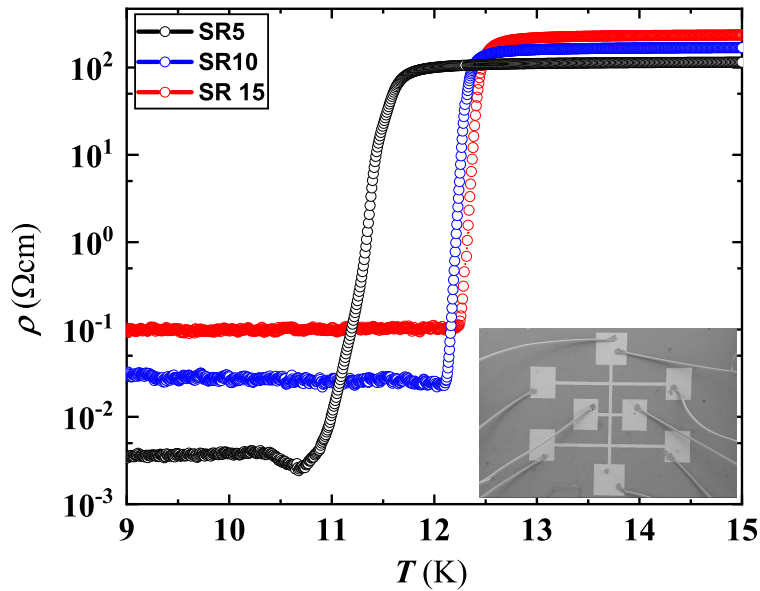}
\caption{NbN resistivity behaviour around $T_c$, for the set deposited on the Al$_2$O$_3$ r-cut substrate (see Table~\ref{tab1}). Inset: Scanning electron microscopy of a typical Hall bar shaped film with Al wires bonded to the sample pads. Current carrying contacts are located on top and bottom, along the vertical line, while a couple of lateral contacts, from the same side, is used to detect the voltage drop.}
\label{rho_vs_T}
\end{figure}
The $\rho(T)$ curves around $T_c$ for the set deposited on the Al$_2$O$_3$ r-cut substrate, is reproduced in Figure~\ref{rho_vs_T}. The SC transition lowers, with the decrease of the film thickness, becoming substantial passing from $d = 10$ nm to $d = 5$ nm.
Normal state resistivity doesn't show, in general, a clear correlation with the value of $d$, appearing even reversed with respect to the $d$ value, for the set deposited on sapphire r-cut (Figure~\ref{rho_vs_T}). However, normal state resistivity variation is confined within a half order of magnitude for $5 \leq d\leq 15$ nm. The behaviour of $\rho(T)$ for SR5, is analyzed in detail in the next section.\\ 
Generally, higher $T_c$ together with narrow SC transition width, $\Delta T_c$, have been measured on MgO substrates above $d = 10$ nm (figure~\ref{T_c} and Table~\ref{tab1}) while a consistent worsening of these parameters occur on SiO$_2$.\footnote{This behaviour has been detected also in film of higher thicknesses, not reported in the present work.} 
Intermediate values of both $T_c$ and $\Delta T_c$ have been measured on Al$_2$O$_3$ substrates, being $T_c$ on the c-cut type slightly higher at $d>5$ nm (figure~\ref{T_c}). 
Compared to NbN films deposited on Al$_2$O$_3$ r-cut, those on c-cut type show a wider range of $T_c$ variation with $d$, while the tendency to the narrowing of $\Delta T_c$ (Table~\ref{Table_Table 1}) with the increase of $d$ denotes an improvement of the film quality, in agreement with findings reported by other groups \cite{joshi2018superconducting,hazra2016superconducting}.\\
\begin{figure}[H]
\centering
\includegraphics[width= 12 cm]{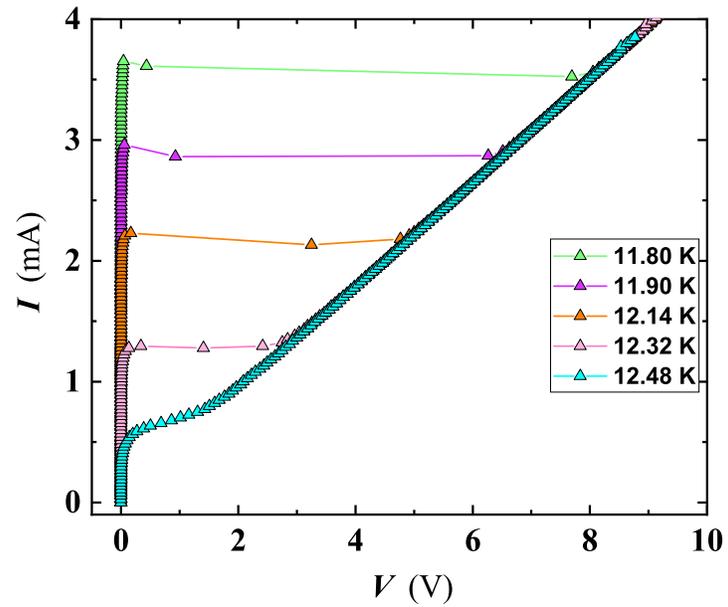}
\caption{I-V curves of the film SC15, measured at several $T$ close to the SC transition of $T_c$ = 12.73 K. approaching $T_c$, the amplitude of the  hysteresis between the sweep-up and sweep-down of the current narrows.}
\label{I-V}
\end{figure} 
While the general trend of $T_c$ as a function of $d$ has been reported in the literature by several groups \cite{Kang2011,Smirnov2018}, $T_c$ spreading depends also on the crystal structure \cite{Senapati2006}, deposition technique, the partial pressure of nitrogen used during the film fabrication process, etc., making difficult a direct comparison of results measured by different groups \cite{Kang2011,Smirnov2018,Sidorova2020,Chockalingam2008}.
Anyway, taking into account NbN films deposited on Al$_2$O$_3$ r-cut substrates, our $T_c$ values while appearing scattered, at $d\leq 10$ nm, follow a trend similar to that found by Soldatenkova et {\it al.} on the same substrate type (see Figure~\ref{T_c}) \cite{soldatenkova2021normal}. Scattering of $T_c$ in NbN films reflects inhomogeneity issues that are characteristic of this superconducting system \cite{Venditti2019}.\\ 
The NbN film properties have been studied further, carrying out current-voltage characteristics. 
Temperature dependent I-V curves exhibit hysteresis and a well defined transition from the superconducting to the normal state for $d > 10$ nm (figure~\ref{I-V}), while at 5 nm and 10 nm of thickness, several NbN films have evidenced the presence of small steps, along the SC transition branch of the I-V curves, better detailed in the next section. For these films, we have assumed as critical current, $I_c$, the value at which the first step occurs.  
The value of the critical current, $I_c$, progressively reduces with the raise of $T$, while smoothing of the transition occurs approaching $T_c$. 
The temperature dependence of the superconducting critical current density, $J_c(T)$, for the set deposited on Al$_2$O$_3$ r-cut substrate (see Figure~\ref{Jc_vs_T}), has been derived following the criterion and procedure reported in the ref.~\cite{Pinto2018}. The $J_c$ values at 0 K (i.e., $J_{c0}$), has been extracted by a least squares fit using the Ginzburg-Landau equation \cite{Pinto2018}.\\
The thickness dependence of $J_{c0}$ appears related to the substrate types exhibiting a bell shaped behaviour on both type of sapphire substrates, while it continues to rise with film thickness on MgO (inset of figure~\ref{Jc_vs_T} and Table~\ref{tab1}).
It's worthwhile noting the drop of $J_{c0}$ at $d = 5$ nm, up to about one order of magnitude, becoming practically independent on the substrate type, while little differences in $J_{c0}$, start to appear from $d = 10$ nm (inset of figure~\ref{Jc_vs_T}). 
The $J_{c0}$ values found in our films are similar to those found in thin films of NbN \cite{il2005critical} while the bell shaped behaviour has been detected also in Nb \cite{Pinto2018}.

\subsection{Berezinskii-Kosterlitz-Thouless transition}
\label{sec:RT}
\begin{figure}[H]
\centering
\includegraphics[width= 11 cm]{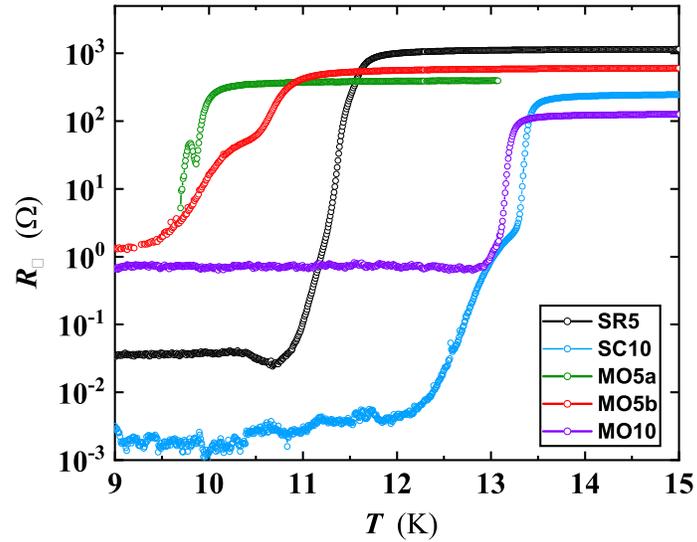}
\caption{Temperature dependent sheet resistance of ultrathin NbN films (5 and 10 nm of thickness), deposited on different substrate types, exhibiting quantum effects in 2D. The difference in the curve behaviour of MO5a and MO5b (belonging to the same deposition run), are related to a specific choice of process parameters during the Hall bar fabrication.} 
\label{BKT_transition}
\end{figure}

\begin{figure}[H]
\centering
\includegraphics[width= 11 cm]{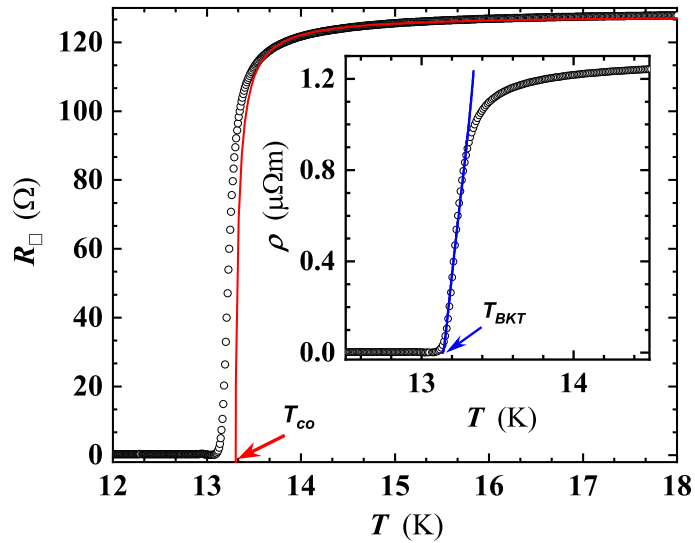}
\caption{Temperature dependent sheet resistance around the superconducting transition temperature for MO10 (see Table~\ref{tab2}). The red line is the least-squares fit by using the Eq.~\ref{R_square}. The value of $T_{c0} = 13.31$ K (red arrow) is the intercept of the red line with the x-axis. Inset: least-squares fitting of the $\rho(T)$ curve of MO10 by the Eq.~\ref{rho_VAP}. The intercept with the x-axis gives the value of $T_{BKT}$ (13.06 K, blue arrow).}
\label{R_sheet}
\end{figure}

\begin{figure}[H]
\centering
\includegraphics[width= 12 cm]{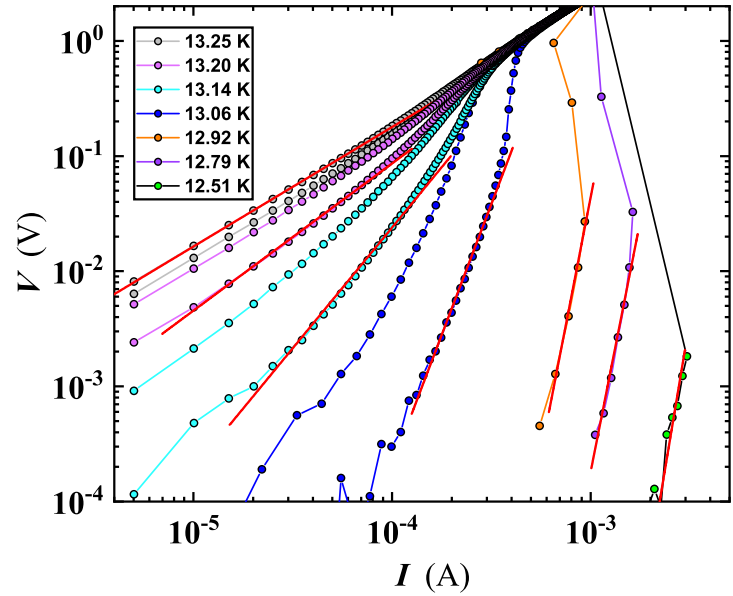}
\caption{Voltage-current characteristics for the film MO10. Experimental data of the sweep-up curve have been fitted by a power law function to extract the value of $\alpha$ (see Eq.s~\ref{power_law} \&~\ref{alpha} and Table~\ref{tab2}).}
\label{I_V}
\end{figure}

\begin{figure}[H]
\centering
\includegraphics[width= 12 cm]{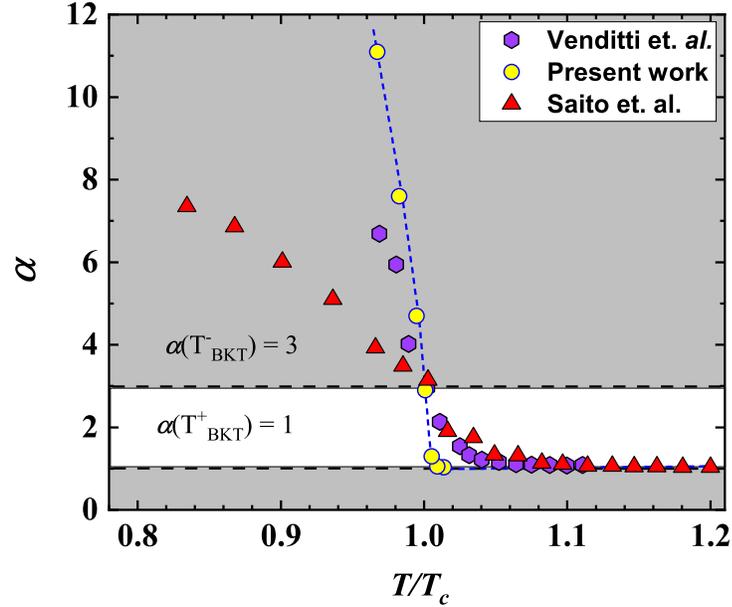}
\caption{Temperature dependence of $\alpha$ for MO10, derived from the power-law fitting of the V-I curves plotted in the figure~\ref{I_V}. A jump in the value of $\alpha$ from $\simeq$1 to $\simeq$3 (white region) has been detected at $T_{BKT}$ corresponding to $T/T_c = 0.989$. For comparison, the $\alpha$ values for a 3 nm thick NbN film from the work of Venditti et {\it al.} \cite{Venditti2019} and for MoS$s_2$ from the work of Saito  {\it al.}  \cite{saito2020dynamical} have been added to the data. Broken line is a guide for the eyes. The temperature has been normalized to $T_{BKT}$ for our and Saito et {\it al.} data. Venditti et {\it al.} $\alpha$ values have been derived by digitization of figure (2e) in the ref. \cite{Venditti2019}.}
\label{alpha_T}
\end{figure}

Experimental observation of a BKT transition in 2D systems is generally challenging due to the constraints on the film size in relation to the characteristic lengths of the superconductor. 
Concerning the Pearl length, the condition $\Lambda > w$, must be fulfilled, $w$ being a film dimension. Hence, assuming for NbN films a value of $0.5 \lesssim \lambda \lesssim 0.4$ $\mu$m for thicknesses $5 \leq d \leq 15$ nm, respectively \cite{Kamlapure2010}, we get values of $\Lambda \approx$100 $\mu$m, $\Lambda \approx$40 $\mu$m and $\Lambda \approx$20 $\mu$m for $d = 5$ nm, 10 nm and 15 nm, respectively. These values must be compared to the maximum physical dimension of the system, that for our films coincides the width of the Hall bar ($w = 50$ $\mu$m)\footnote{For the Hall bar width of 10 $\mu$m the condition is satisfied at any of the thicknesses here taken into account.}.
Hence, while the 10 nm thick sample can be considered border line, the 15 nm thick one appears out of range. Regarding the condition $d \ll \xi$ it's worthwhile noting that being $\xi \approx$ 4 nm \cite{Chockalingam2008} only NbN nanofilms, 5 nm thick, appear to be good candidates to exhibit a well defined BKT transition.\\
Experimentally, we have investigated the signature of a BKT superconducting transition in $\rho(T)$ curves and/or in I-V characteristics measured at several fixed $T$s \cite{Koushik2013,Bartolf2010,Venditti2019,Sidorova2020}.\\ 
We have extracted experimentally the $T_c$ value in zero magnetic field, $T_{c0}$, and the sheet resistance in the metallic normal state, $R_{\Box N}$, just above $T_c$, considering a Cooper-pair fluctuation model for a 2D superconducting system developed by Aslamazov and Larkin (AL) \cite{AL1968} and Maki and Thompson (MT) \cite{Maki1968,Thompson1970}. The two parameters have been evaluated by a least-squares fitting of the experimental $R_{\Box}(T)$ curves  (Figure~\ref{BKT_transition}), in a $T$ range from $T_c$ to $\approx 15$ K, by using the relation \cite{Bartolf2010,Sidorova2020}:
\begin{equation}
\label{R_square}
R_{\Box}(T) = \frac{R_{\Box N}}{1+ R_{\Box N}\frac{\gamma}{16}\frac {e^2}{\hslash}(\frac{T_c}{T-T_c})}
\end{equation}
where: $\gamma$ is a numerical factor, $\hslash$ is the reduced Planck constant, $e$ is the electron charge and $T_c$ is here intended as the BCS mean-field transition temperature. Fitting has been carried out satisfying the condition $ln(T/T_c) \ll 1$. The result of the $R_{\Box}(T)$ fitting of film MO10, is reproduced in figure~\ref{R_sheet}. Similar findings have been found also for MO5a and MO5b. 
\begin{table}[H] 
\caption{Berezinskii-Kosterlitz-Thouless (BKT) parameters derived by the analysis of the resistivity and I-V characteristics curves of some of the thinner NbN films. Columns heading, from left: sample acronym\textsuperscript{*}; BKT temperature derived by $\rho(T)$ fitting with Eq.~\ref{rho_VAP}; SC transition temperature at $B$ = 0 (see the text); normal state sheet resistance at 20 K; $\gamma$ value (see the text and Eq.~\ref{R_square}); VAP polarizability.}
\label{tab2}
\newcolumntype{C}{>{\centering\arraybackslash}X}
\begin{tabularx}{\textwidth}{CCCCCCC}
\toprule
\text{Sample}	& \text{$T_{BKT}$ (K)}   & \text{$T_{c0}$ (K)}  & \text{$R_{\Box N}$ ($\Omega$)}  & \text{$\gamma$}  & \text{$\epsilon$}\\
\midrule
MO5a	& 9.75    & 9.94    & 502.33    & 0.97      & 10.28  \\
MO5b	& 10.36     & 10.68    & 620    & 1.50    & 11.49  \\
MO10	& 13.13     & 13.31    & 129.8    & 1.56    & 25.52  \\
\bottomrule
\end{tabularx}
\noindent{\footnotesize{\textsuperscript{*} For meaning of sample acronym see the caption of Table~\ref{tab1}.}}
\end{table}
Fitted and experimental $R_{\Box N}$ values agree within $2\div3$\%, while $T_c$ and $\gamma$ exhibit a dependence on $d$ and the substrate type, this last particularly effective on $T_c$ values for NbN films of 5 nm of thickness (see Figures~\ref{BKT_transition}).\\ 
In detail, $T_c$ values derived by the fit are close, though systematically lower, than those extracted by the analysis of the SC transition branch of the $\rho(T)$ curves (see the Table~\ref{tab1}), while the values of $\gamma$, ranging from $\approx1$ to $\approx2$ (Table~\ref{tab2}), are in agreement with those reported for NbN films with $d < 10$ nm \cite{Bartolf2010,Sidorova2020} and, in general, comparable with those reported in literature for different SC materials.\\
Aimed to check the possible BKT-like transition, we have carried out further analyses considering that thermal fluctuations occurring in ultra thin films can excite pairs of vortices each consisting in a single vortex having supercurrents circulating in opposite directions, then leading to the bound  vortex anti vortex pair state \cite{Bartolf2010}. These VAP pairs leads to the specific signature of a BKT-like transition, consisting a jump in the superconducting stiffness, $J_s$, from a finite value, below $T_c$, to zero above it. In that case, a non linear dependence exists in the I-V characteristics near $T_c$, since large enough currents may unbind VAPs. Hence, due to this effect, a voltage is generated, depending on the equilibrium density of the free vortices, scaling with the sourced current according to a power law, with an exponent proportional to $J_s$:
\begin{equation}
\label{power_law}
V \propto I^\alpha(T)
\end{equation}
\begin{equation}
\label{alpha}
\alpha(T) = 1+\pi\frac{J_{s}(T)}{T}
\end{equation}
At the BKT transition, the I-V exponent jumps from $\alpha(T^-_{BKT}$) = 3 to $\alpha(T^+_{BKT})  = 1$, where the unity value signals the metallic ohmic behavior in the normal state of the I-V characteristics. 
We have extracted the value of $\alpha$ for the MO10 film, by using Eq.~\ref{power_law} carrying out a least squares fitting of the voltage-current curves, measured at several $T$ (figure~\ref{I_V}), detecting the universal jump from $\simeq$1 to $\simeq$3 at $T_{BKT}$ (figure~\ref{alpha_T}). Our findings evidence a steeper transition (see figure~\ref{alpha_T}) resulting in agreement with data of Venditti et {\it al.} for a thin film of NbN \cite{Venditti2019}. The behaviour here reported for the NbN system has been also found by Saito et {\it al.} for a completely different system as the MoS$_2$ \cite{saito2020dynamical} (figure~\ref{alpha_T}) which validates the universal jump in the superfluid density for determining the BKT transition.
Based on theoretical studies, below $T_{BKT}$, all VAPs are bound, while approaching $T_{BKT}$, thermal fluctuations begin to break VAPs and, under thermodynamic equilibrium, VAPs and single vortices will coexist \cite{Bartolf2010,bell2007one,benfatto2009broadening}. However, due to the sourced current, single vortices will experience a Lorentz force (neglecting vortex pinning) causing the appearance of a finite voltage drop. The resulting film resistivity, in the temperature range $T_{BKT} < T<T_{c0}$, can be described by the relation:
\begin{equation}
\rho(T) = a exp\left (-2 \sqrt{b \frac{T_{c0}-T}{T-T_{BKT}}}\right)
\label{rho_VAP}
\end{equation}
where $a$, $b$, are fitting parameters related to the SC material. 
The $T_{BKT}$ value derived by the least squares fitting with Eq.~\ref{rho_VAP} of the $\rho(T)$ curve of film MO10 is shown in the inset of figure~\ref{R_sheet}. The fitting procedure has been extended to other thin films, for which no such sign of BKT transition has been detected.
It is interesting to estimate the polarizability, $\epsilon_{BKT}$, of a VAP at the BKT-like vortex phase transition in presence of other VAPs, by using the relation \cite{Bartolf2010}:
\begin{equation}
\label{TBKT}
\frac {T_{BKT}}{T_{c0}}= \frac{1}{1+0.173\epsilon_{BKT}R_{\Box N}\frac {2e^2}{\pi h}}
\end{equation}
The Eq.~\ref{TBKT} has been applied to NbN films deposited on different substrates, resulting in values of $\epsilon$ (Table~\ref{tab2}) in close agreement with those found in ref. \cite{Bartolf2010} for NbN film 5 nm thick and also successfully crosschecked with the universal relation $k_BT_{BKT} = AT_{BKT}/4\epsilon_{BKT}$ for topological 2D phase transitions, shown by Nelson and Kosterlitz \cite{yamashita2011origin,nelson1977universal}.\\
\subsection{Phase slips}
\label{sec:PS}
\begin{figure}[H]
\centering
\includegraphics[width= 11 cm]{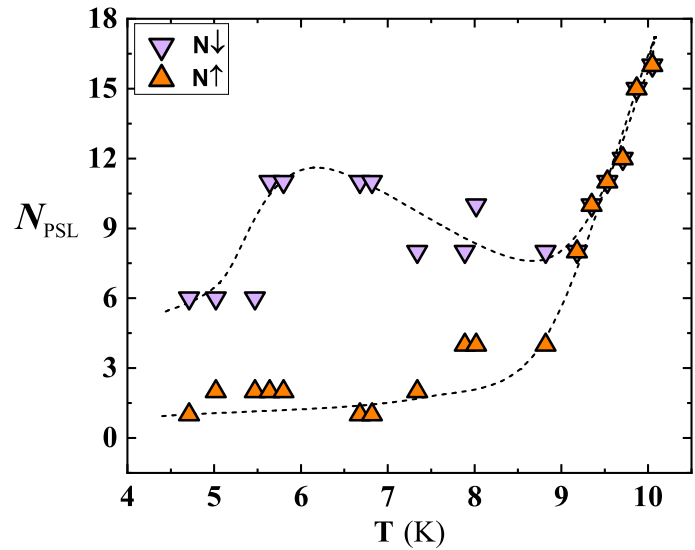}
\caption{Number of resistive states for SR5, calculated from the I-V curves, during the current sweep-up ($N^\uparrow$) and sweep-down ($N^\downarrow$). Close $T \simeq 9$ K, $N^\uparrow$ and $N^\downarrow$ suddenly converge, assuming the same value at $T\geq 9$ K. Broken lines are guides for eyes.}
\label{NPSL}
\end{figure}

\begin{figure}[H]
\centering
\includegraphics[width= 11 cm]{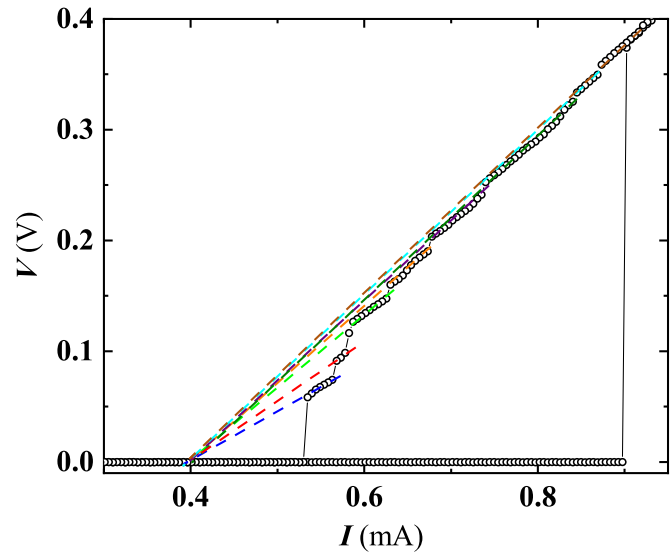}
\caption{V-I curves at few selected temperatures for SR5, showing the convergence of all resistive tails towards a well defined value of the excess current of {$I_{ex} = 0.38$ mA}.}
\label{I-V_intercept}
\end{figure}

In addition to the BKT transition, interesting outcomes have been found in our NbN films, carrying out I-V characteristics by the pulse mode technique (see the Materials and Methods section). In detail, for two of our thinnest films (SR5 and MO5b), we have observed the emergence of resistive states, as tailing-like features in I-V curves and a double transition in the $\rho(T)$ curves. These findings have been interpreted as possible signatures of phase slip events, arising due to a discontinuous jump by integer multiple of $2\pi$ in the phase of the order parameter of the superconducting state, typically existing in quasi-1D systems, as nanowires and nanorings.\\ 
Nevertheless investigated NbN films are a 2D system which are granular in nature, the presence of disorder can lead to a weak localization and inhomogeneous effects, resulting in the appearance of a 1D-like features such as PS events \cite{bell2007one,gajar2019substrate,chu2004phase}.\\
Taking into account the above mentioned scenario, such systems are prone to form an array of continuous conductive paths, having an effective dimension much smaller than the physical dimension of the system. In that case, a region equivalent to the coherence length can create a Josephson-like junction, which shows a PS barrier proportional to the area of quasi-1D SC system. Under these circumstances, Cooper pairs will cross the free energy barrier and the relative phase will jump by 2$\pi$, resulting in a measurable voltage drop. This drop, will cause a detectable resistance change at $T < T_c$, giving rise to PS events driven by thermal and quantum activation \cite{chu2004phase}. 
To confirm the possible presence of PS in our films, we have taken into account the theory of Langer, Ambegaokar, McCumber, and Halperin (LAMH), accounting for thermally activated phase  slips (TAPS) detected as an effective resistance change, related to the time evolution of the superconducting phase:
\begin{equation}
R_{TAPS} = \frac{\pi \hbar^2 \Omega_{TAPS}}{2e^2K_BT} exp\left(\frac{-\Delta F}{K_BT}\right)
\label{R_TAPS}
\end{equation}
where $\Delta F$ is the energy barrier to Cooper-pair crossing, while the other quantities have the known meaning. Here, $\Omega_{TAPS}$ is the attempt frequency defined as:
\begin{equation}
\Omega_{TAPS} = \frac{L}{\xi}\left(\frac{\Delta F}{K_BT}\right)\frac{1}{\tau_{GL}}   
\label{Omega_TAPS}
\end{equation}
where $\tau_{GL}$ is the relaxation time in the time-independent Ginzburg-Landau equation and $\Delta F$ is defined as:
\begin{equation}
\Delta F = 0.83 K_{B}T_c \frac{L\beta w d R_q}{\rho_n \xi_0}
\label{Energy_barrier}
\end{equation}
where: $\beta$ is a fitting parameter; $w$ is the Hall bar width; $d$ the film thickness; $R_q = \hbar/2e^2 \simeq 6.4$ k$\Omega$ is the quantum of resistance; $\rho_n$ the normal state resistivity upon SC transition.
\begin{figure}[H]
\centering
\includegraphics[width= 11 cm]{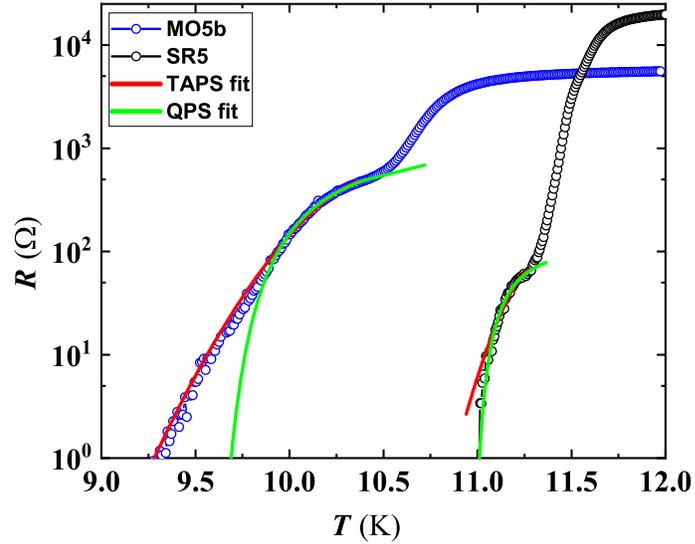}
\caption{Temperature dependence of resistance for two 5 nm thick NbN films. Both curves show tailing-like features below about 10.5 K and a 11.5 K for MO5b and SR5, respectively. Branches of the $\rho(T)$ curves exhibiting the transition at the lower $T$ range, have been fitted by using LAMH Eq.(\ref{R_TAPS}) (red lines) and by Eq. (\ref{GIO}) (green lines).}
\label{tpsfit}
\end{figure}
The LAMH theory was originally developed for very long wires, thinner than $\xi$ of the SC material. However, our investigated NbN films, have a thickness comparable to $\xi$ while $w$ and $L$ (i.e., width and the length of the Hall bar, respectively) are orders of magnitude greater then $\xi$.\\
In order to check the presence of TAPS in SR5 and MO5b, least squares fits have been carried out by Eq.~\ref{R_TAPS}, leaving $T_c$ and $\beta$ as fitting parameters. For MO5b, good agreement with the LAMH theory has been found, with $T_c$ and $\beta$ values comparable to those found in the ref. \cite{chu2004phase} (see figure~\ref{tpsfit}). 
Moreover, the $T_c$ value obtained by the LAMH fit coincides with that derived from resistance curve analysis (see Table\ref{tab1}), confirming that fluctuation effects are caused by PS and have a thermal origin.\\ 
On the other hand, for SR5 at lower $T$, fitting by LAMH theory failed in the first steeper branch of the $\rho(T)$ curve (see figure~\ref{tpsfit}) suggesting the possibility of a different fluctuation effect present in the same $T$ range. 
To confirm our hypothesis, we have explored the possible contribution by QPS, emerging from the quantum tunneling of the order parameter through the same free energy barrier as in TAPS, which are supposed to dominate at lower $T$. The dynamics of the order parameter in the quantum fluctuations was first reported by Giordano \cite{giordano1991superconductivity}, suggesting a mechanism similar to TAPS, except that appropriate energy scale $K_BT$ is replaced by $\hbar /\tau_{GL}$, resulting in the equation: 
\begin{equation}
R_{QPS} = B\frac{\pi \hbar^2 \Omega_{QPS}}{2e^2 \frac{\hbar}{\tau_{GL}}}exp\left(-a\frac{\Delta F \tau_{GL}}{\hbar}\right)  
\label{GIO}
\end{equation}
where $B$ and $a$ are numerical factors of $\approx1$ and $\Omega_{QPS}$ is defined as:
\begin{equation}
\Omega_{QPS} = \frac{L}{\xi}\left(\frac{\Delta F}{\frac{\hbar}{\tau_{GL}}} \right)\frac{1}{\tau_{GL}}   
\label{Omega_QPS}
\end{equation}
The result of the fitting by Eq.~\ref{GIO} is in agreement with the theoretical predictions (see figure~\ref{tpsfit}) for SR5, while on the contrary, a progressive deviation from the QPS equation (\ref{GIO}) is evident at lower $T$ for MO5b.\\ 
The excellent agreement of our experimental results of $\rho(T)$ with the above mentioned LAMH and QPS models, suggest the existence of a nano-conducting path (NCP) having a lateral size comparable to $\xi$. To estimate this size, we have used the equation from the model developed by Joshi et {\it al.} \cite{joshi2020dissipation}:
\begin{equation}
d_{NCP}= \sqrt{\left(\frac{12 \rho_{RT} C \xi_0}{1.76 \sqrt 2 \pi R_q} \right)}
\label{NCP}
\end{equation}
where $\xi_0 \simeq 4$ nm, $C = 8$ (value typically used for quantum systems) \cite{Bezryadin2008} and $\rho_{RT}$ is the resistivity value at RT. Calculated NCP values by using Eq.~\ref{NCP} are 3.2 nm for SR5 and 7.5 nm for MO5b,  comparable to those found for NbN nanostructures in the ref. \cite{joshi2020dissipation} are in agreement with our experimental results. In fact, for SR5, the condition $d_{NCP} < \xi$ (i.e., $\xi \approx4$ nm), can be considered as the origin of quantum tunneling, resulting in the formation of QPS detected in our film. On the contrary, for MO5b, being $d_{NCP}> \xi$ TAPS lines will result, as indeed experimentally observed.\\
An interesting outcome derived by a detailed analysis of I-V characteristics carried out on SR5 and MO5b confirms the existence of PS events. 
\begin{figure}[H]
\centering
\includegraphics[width= 12 cm]{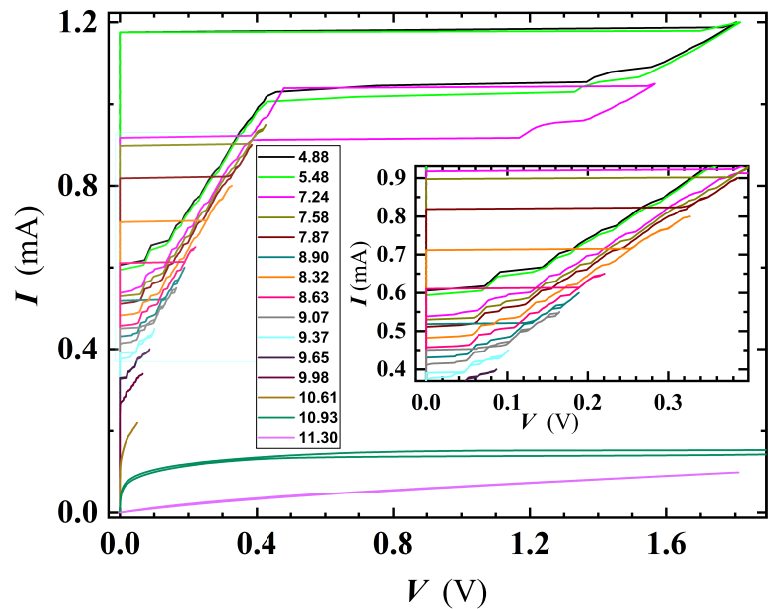}
\caption{I-V characteristics for the SR5 film, carried out at several $T$. Both up and down current sweeps, evidence the presence of slanted steps due to the occurrence of intermediate resistive regimes before the complete transition to the normal state. Inset: magnification of the central part of the plot.} 
\label{PSL}
\end{figure}
Figure~\ref{PSL}, shows a family of I-V curves, measured at different $T$ for SR5, where multiple slanted steps, having resistive tailing-like features have been detected. The dynamic resistance of these tails rises, at increasing $V$ (e.g., along the I-V sweep-up direction, see figure~\ref{PSL}), while the current axis intercept (extending tails slope) occurs to a common current value of $I_{ex}\simeq$ 0.38 mA, defined as excess current (see Figure~\ref{I-V_intercept}). This finding can be considered as a proof that detected resistive states are originated by PS \cite{Sivakov2003}.\\
Further analysis of the I-V curves, has evidenced an interesting temperature dependence of the number of the resistive states appearing in the sweep-up ($N^\uparrow$) and sweep-down ($N^\downarrow$) branches of the I-V curves (figure~\ref{NPSL}). The $N^\uparrow$ and $N^\downarrow$ keep an almost constant value up to $T \approx T_c/2$, and then they start to differ. In detail, at $T \gtrsim T_c/2$, $N^\downarrow$ jumps to higher values, and then it decreases till about 9 K. On the contrary, $N^\uparrow$ continue to gradually increase till 9 K. Onward 9 K, both curves are perfectly overlapped and continuously rise with $T$ (figure~\ref{NPSL}). The behaviour exhibited by $N^\uparrow$ and $N^\downarrow$ allows to derive additional observations about fluctuation effects involved in our investigated film. Specifically, the unequal distribution of PS in $N^\uparrow$ and $N^\downarrow$ suggests that different mechanisms are contributing in three distinct $T$ ranges. Below $T_c/2$, the emergence of phase slip events is driven by quantum tunneling. In the intermediate $T$ range ($T_c/2<T<9$ K), the behaviour of $N^\uparrow$ and $N^\downarrow$ can be explained by the competition of QPS and TAPS, the former tending to decrease approaching 9 K. Finally, above 9 K, only TAPS contribute in the observation of PS events (figure~\ref{NPSL}). It is interesting to note that in the region where QPS is present, $N^\uparrow$ and $N^\downarrow$ diverge. This effect can be explained by a presumably different tunneling route followed by the system, during the sweep up and sweep down of the sourced current in the quantum regime. In the intermediate $T$ range, the system self organize in order to converge towards a condition dominated by TAPS, then there is a decrease in $N^\downarrow$ with an increase $N^\uparrow$. Finally, approaching $T_c$, thermal fluctuations dominates over QPS and the entire system is driven by the electrons. Since the electrons tends to track the same path while going sweep-up and sweep-down \cite{kumar2021switching}, the number of PS becomes equal followed by a total increase in the number of PSL, due to enhanced fluctuation effects near $T_c$. \\
In addition to TAPS, BKT-like features have been also observed in MO5b, in the $T$ range where PS disappear and the system still remains in the SC state. Moreover, in the same film, BKT has been confirmed by the scaling behaviour of Halperin-Nelson equation above $T_{BKT}$ (as explained in the BKT section). Interestingly, from our I-V characteristics, the TAPS completely disappears at 9.9 K, way before reaching $T_c$ (11.02 K, see Table~\ref{tab2}) and this PS suppression gives rise to BKT-like transition, i.e. an effect typical of a 2D system, which involves breaking of VAP. For $T>T_{BKT}$, the unbinding of VAP will result into the universal jump in $\alpha$, as already detected in our investigation of the I-V characteristics of MO10 (see figure~\ref{I_V}). The derived parameters for this film (see Table~\ref{tab2}) are in good agreement with the theory. We believe that inhomogeneity of the NbN system can contribute to the origin of these experimental findings. However, our results suggests that the level of inhomogeneity in the studied system is just sufficient to create a thermal fluctuation which results in emergence of TAPS but not too strong to destroy the superconductivity of the film. The detection of BKT in the same film seems to confirm this hypothesis, since higher inhomogeneity would have been destroyed the BKT effect. The coexistence of BKT and TAPS in the same system, the former typical of a 2D system, the latter of a quasi 1D system, suggests that the internal structure of our investigated films is on the boundary line of a 2D-1D dimensional crossover.
\begin{figure}[H]
\centering
\includegraphics[width= 11 cm]{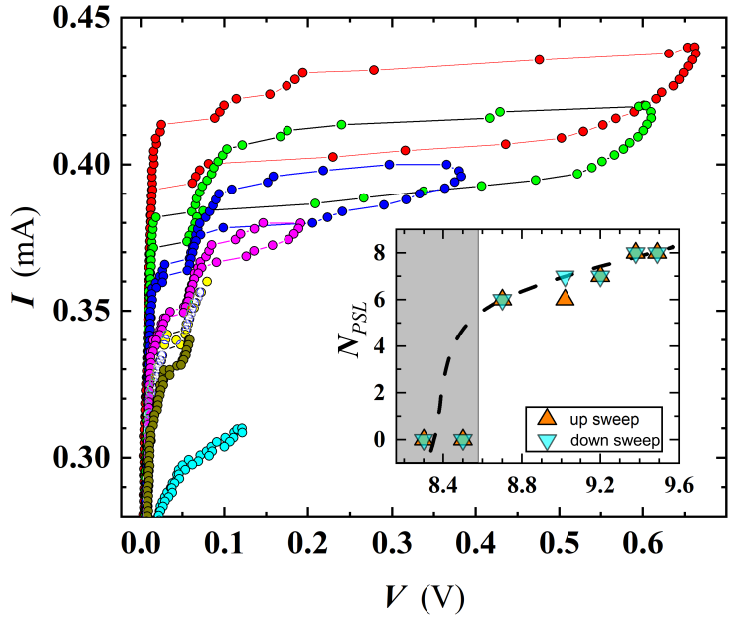}
\caption{I-V characteristics of MO5b, showing the occurrence of steps in the curves close to $T_c$, progressively disappearing approaching $T_c$. The $T$ values for the shown curves are: 8.70 K (red), 9.02 (green), 9.20 K (blue), 9.38 K (magenta), 9.49 K (yellow), 9.60 K (olive), 9.98 K (cyan). Inset: number of TAPS extracted from I-V curves. The grey rectangle defines the $T$ range where no PSL have been detected. }
\label{MO5}
\end{figure} 
Similar findings have been measured at 5 nm of thickness, for NbN films deposited on other type of substrates. However, on SiO$_2$, resistive tails appear less pronounced due to the larger width of the Hall bar (see figure~\ref{I-V_PSL_SiO2}). 
\begin{figure}[H]
\centering
\includegraphics[width= 12 cm]{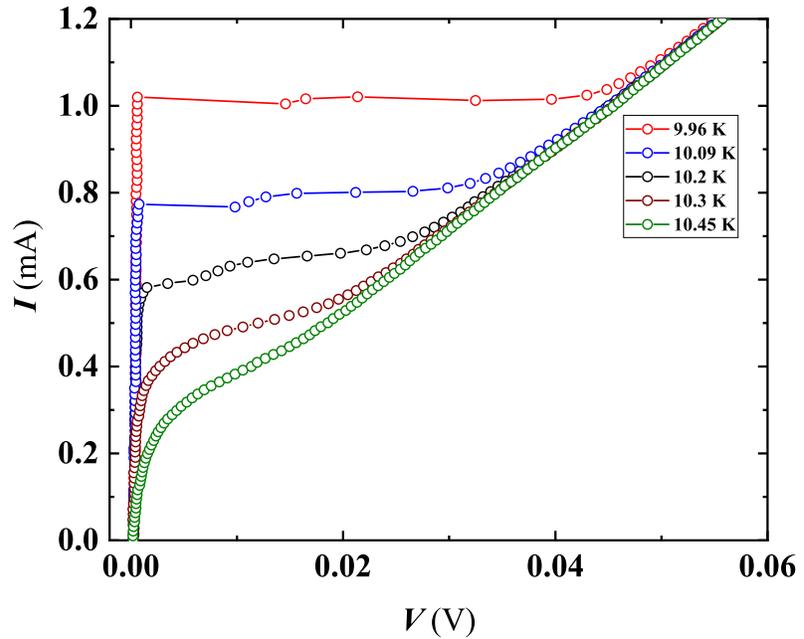}
\caption{I-V characteristics for SO5, at selected $T$ values close to $T_c$. Less marked resistive tails are now visible, whose number is reduced if compared to those detected in the film SR5 (see the figure~\ref{PSL}).}
\label{I-V_PSL_SiO2}
\end{figure} 
\section{Conclusions}
In summary, our detailed study of superconducting properties of NbN ultra-thin films has shown several interesting features associated with complex phase fluctuations of the superconducting order parameter. The extensive investigation of resistivity and I-V curves has shown a well defined BKT transition to the superconducting state with quasi-long-range order, characteristic of 2D systems. In addition to the BKT physics, we have also detected and characterized phase slip events (both quantum and thermal) which are typical of quasi-1D superconducting systems. Both effects demand a careful and fine tuning of the experimental set-up and material system.\\    
To perform the analysis of the BKT transition, we have used the established theoretical framework based on the Cooper pair fluctuation model. We found that the linear in $T$ dependence of the resistivity, well above $T_c$, is only compatible with a 2D fluctuation-conductivity and is incompatible with the corresponding predictions for 1D and 3D systems available in the literature \cite{Larkin2003}, confirming the 2D dimensionality of our NbN films. Our findings of polarizability values of VAP at the BKT transition are in good agreement with the Nelson and Kosterlitz universal relation in two different NbN films from the measured set. In one case, the polarizability has been found almost twice the expected value. This fact has been explained as evidence of the dimensional crossover, from 3D to 2D, occurring at 10 nm, since no BKT transition has been detected at 15 nm of film thickness. A further confirm has been gained by the $\alpha$ exponent, that we have extracted from the power-law fitting of the I-V curves, exhibiting a steep transition, from 1 to 3, close to $T_{BKT}$, in agreement with the theoretical predictions.\\ 
Regarding the PS events of the order parameter, we have detected in one sample the presence of quantum and thermal phase slips, in different temperature regimes. Quantum phase slips show features that depend on the way quantum tunneling route is undertaken by the system. 
These outcomes have been explained by the presence of inhomogeneity and granularity in the NbN system. In fact, it is well established experimentally that the microscopic structure of NbN is a collection of randomly distributed nanoscale-sized SC grains, in close contact with each other (Josephson-like system) and separated by grain boundaries. We confirmed by a careful analysis of film resistivity the presence of nano-conductive path, making our NbN films equivalent to a quasi-1D system, which explains the presence of PS events.
We have been able to investigate a specific feature of these phase slip events as the number of PS occurring during the up-sweep and down-sweep of the sourced current. The analysis shows the uneven distribution of PS in the quantum regime, converging to the same value in the temperature regime dominated by thermal fluctuations. 
Finally, our study has evidenced the co-existence of BKT and PS phenomena in the same NbN nanofilm, that to our best knowledge has not been reported so far in a superconducting system having the physical size comparable to our films. Considering that BKT and PS events belong to two different dimensionality systems, that emerge separately in superconductors, this means that we have successfully addressed a dimensional crossover, from 2D to quasi-1D, in the same NbN system. 
A relevant role in these results has been played by the granular nature of NbN films. Our experimental findings motivate for in depth study of superconducting NbN nanofilms as a tunable platform to generate and control novel quantum phenomena exploitable for quantum technologies.
\vspace{6pt} 



\authorcontributions{Conceptualization, M.S, A.P., N.P. M.S.and R.S.; Film deposition, M.Singh and R.K.R.; Device fabrication, M.S., M.F. and N.DL.;  investigation, M.S., N.P.; formal analysis, M.S.; validation, N.P., S.P.S. and A.P.; writing---original draft preparation, M.S. and N.P.; writing---review and editing, all authors; supervision, N.P.; All authors have read and agreed to the published version of the manuscript.}

\funding{This research received no external funding.}

\dataavailability{Not applicable.} 

\acknowledgments{We acknowledge the University of Camerino for providing technical and financial support. The School of Science and Technology and the Physics Division are acknowledged for their contribution to the installation of the SEM and the He closed cycle cryostat equipments. We acknowledge fruitful discussions with Sergio Caprara and Luca Dell'Anna.}

\conflictsofinterest{The authors declare no conflict of interest.} 

\begin{tabular}{@{}ll}
SC & Superconducting \\
BKT & Berezinskii Kosterlitz Thouless\\
PS & Phase slip\\
TAPS & Thermally activated phase slips\\
QPS & Quantum phase slips\\
NCP & Nano conducting path
\end{tabular}

\begin{adjustwidth}{-\extralength}{0cm}

\reftitle{References}



\bibliography{Manuscript}


%


\end{adjustwidth}
\end{document}